\documentclass[aps,pra,groupedaddress,twocolumn]{revtex4-2}

\usepackage{graphicx}
\usepackage{dcolumn}
\usepackage{amsmath}
\usepackage{amssymb}
\usepackage{bm}

\begin{document}



\title{Atomic magnetometry based on the ground-state Hanle effect\\in an elliptically polarized light wave}


\author{Denis Brazhnikov}
\email[Corresponding author: ]{x-kvant@mail.ru}
\author{Anton Makarov}
\author{Katerina Kozlova}
\author{Andrey Goncharov}
\affiliation{Institute of Laser Physics SB RAS, 15B Lavrentyev Avenue, Novosibirsk 630090, Russia}

\date{\today}

\begin{abstract}
We investigate the ground-state Hanle effect in alkali-metal vapor irradiating by a resonant elliptically polarized light wave. The magneto-optical resonances are observed as a change in the ellipticity parameter of the light wave polarization when scanning the transverse magnetic field near zero. We use a miniature ($\approx\,$$0.125$~cm$^3$) glass cesium vapor cell heated to a relatively low temperature of $\approx\,$$85^\circ$C. Under the current experimental conditions, the sensitivity of magnetic field measurements is limited by a technical noise, reaching $180$~fT/$\surd$Hz in a $200$~Hz bandwidth. The ultimate photon-shot-noise-limited sensitivity of the method is estimated to be $\approx\,$$5$~fT/$\surd$Hz. The proposed scheme is promising for the development of a zero-field atomic magnetometer with reduced heat dissipation of the sensor head and relaxed requirements for magnetic shielding compared to counterparts operating in the spin-exchange relaxation-free regime. These features are of particular value for biomedical applications.
\end{abstract}

\maketitle

\newcommand{\SixJ}[6]{
\begin{Bmatrix}
#1 & #2 & #3\\
#4 & #5 & #6
\end{Bmatrix}
}

\section{\label{sec:1}Introduction}

Atomic magnetometers (AMs) are the most sensitive quantum sensors of magnetic fields \cite{Allred2002,Kominis2003,Budker2007}. Their operation relies on laser spectroscopy techniques applied to atoms, most commonly alkali metals (Rb, Cs, K), as well as helium atoms. AMs enable the solution of a broad range of applied and fundamental problems. For instance, in medicine, AM-based approaches to magnetocardiography (MCG) \cite{Xiao2023,Chen2025} and magnetoencephalography (MEG) \cite{Osborne2018,Boto2022,Alem2023,Fedosov2025} are being actively developed. In biology, AMs allow the detection of plant biomagnetism \cite{Fabricant2021}. In materials science, AMs can be employed for the investigation of nanoparticle properties \cite{Taskova2022} and material tomography \cite{Wickenbrock2013}. In fundamental physics, AMs are utilized, for example, in searches for dark matter particles \cite{Afach2018}. These sensors have also been employed in scientific space missions \cite{Ellmeier2023}.

To date, a variety of atomic magnetometry techniques have emerged, each characterized by distinct operational principles and advantages, with improvements being made all the time \cite{Aleksandrov2009,Fabricant2023}. Among the most sensitive approaches is zero-field level-crossing magnetometry, also referred to as the ground-state Hanle effect (GSHE) \cite{Gawlik1991,Breschi2012}. The GSHE-based technique underpins one-axis \cite{Kominis2003,Alipieva2005,Shah2007}, two-axis \cite{Papoyan2016,LeGal2019}, or three-axis (full vector) \cite{Boto2022,Dong2012,Azizbekyan2017,LeGal2022} magnetic field measurements. Since magneto-optical resonances (MOR) provided by the GSHE manifest in the vicinity of zero magnetic field, achieving high sensitivity of measurements necessitates stringent passive shielding of the sensor head from ambient magnetic fields, which is often complemented by an active compensation system \cite{Holmes2022,Skidchenko2025,Zhang2025,Chen2025}.

For many practical applications, miniaturization of the magnetic sensor is essential, requiring the atomic cell volume to be substantially smaller than $1$~cm$^3$. A reduction in the atom-light interaction region, however, generally comes at the cost of degraded measurement sensitivity. This challenge has been successfully overcome through the spin-exchange relaxation-free (SERF) regime \cite{Happer1977}, which emerges at higher alkali-metal vapor densities. This regime provides a noticeable linewidth narrowing of MOR, thereby boosting sensitivity of miniaturized GSHE-based sensors to levels of $\approx$~$10$$-$$100$~fT$/$$\surd$Hz \cite{Shah2007,Zhang2022}. The SERF regime has become the mainstream in zero-field magnetic-sensor technology, so that it is widely leveraged in state-of-the-art commercial AMs \cite{Osborne2018,Alem2023,Twinleaf}.

For the development of miniaturized magnetic sensors, preference is typically given to single-beam configurations, in which just a single laser beam is employed both for optical pumping (quantum state preparation) and probing the atoms \cite{Osborne2018,Alem2023,Twinleaf,Shah2009,Tang2021,Sebbag2021,Yang2023}. In the most sensitive realizations of such sensors \cite{Shah2009,Tang2021,Sebbag2021,Yang2023}, circular birefringence is induced in the medium by a propagating elliptically polarized wave: the circularly polarized component of the wave acts as the pump wave, while the linearly polarized component serves as the probe one. In these schemes, the laser frequency is detuned far from the absorption line center, typically by tens of GHz. To mitigate adverse effects associated with optical frequency drift, frequency stabilization must be implemented in AMs. However, in the case of laser fields strongly detuned from resonance, this requirement poses a serious technical challenge, demanding nontrivial solutions \cite{Hu2017}. An alternative single-beam scheme that circumvents these limitations employs polarization modulation of the laser field \cite{Petrenko2021}. Nevertheless, this approach is more difficult to realize in miniaturized sensors, especially when dozens of identical sensor heads are required (e.g., as in MEG). Another promising single-beam configuration involves the use of linearly polarized resonant radiation, where an analogue of the SERF effect has been observed \cite{Petrenko2025}. However, an increased vapor cell temperature was required ($120$$^\circ$C).

Despite their advantages, SERF magnetometers exhibit several intrinsic limitations. In particular, activation of the SERF regime requires an increased density of alkali atoms and, consequently, a relatively high temperature of the cell, typically in the range $140$$-$$200^\circ$C \cite{Shah2007,Shah2009,Tang2021,Sebbag2021,Yang2023}. In addition to increased power consumption, such high operating temperatures pose a significant problem when the sensor must be placed in close proximity to the subject under investigation, as in medical or biological studies. For instance, in the context of neonate MEG recordings, this limitation necessitates the placement of a sensor at some distance from the scalp, which in turn diminishes the sensitivity of the measurements \cite{Corvilain2025}.

Another drawback of SERF magnetometers is their limited dynamic range. To achieve optimal sensitivity, these devices require stringent shielding of the sensor head from ambient magnetic fields down to the level of $\sim$~$1$~nT, including suppression of field gradients along all three axes \cite{Zhang2025}. Furthermore, SERF magnetometers typically suffer from a comparatively narrow bandwidth, which is usually confined to the range of $\approx\,$$50$$-$$150$~Hz \cite{Kominis2003,Shah2009,Tang2021,Wang2025}. For these reasons, considerable attention is nowadays paid to the development of miniaturized magnetometer schemes that do not rely on the SERF regime \cite{Petrenko20212,Rushton2023,Bonnet2025}, with the aim of lowering vapor temperature, extending dynamic range and bandwidth of the sensor, while preserving high sensitivity.

In this work, we investigate a GSHE-based magnetometry technique, employing a single elliptically polarized laser beam and a magnetic field scanned in the orthogonal direction to the wave vector (the Voigt configuration). We previously utilized a similar configuration to observe MORs using the Bell-Bloom technique \cite{Makarov2025}. The optical field is tuned to resonance with the Cs D$_1$ line transitions. MORs are detected with the help of polarimetry through changes in the ellipticity parameter of the transmitted beam, which are caused by a circular dichroism of the medium. This detection technique substantially suppresses laser intensity noise, which is one of the dominant noise sources in AMs \cite{Krzyzewski2019}. Moreover, the proposed scheme does not require optical detuning of the laser from resonance with the medium, thereby reducing the undesirable influence of light shifts, which often represent a major challenge in atomic magnetometry \cite{Ma2022,Peng2024}. Finally, tuning the laser frequency to the center of the absorption line almost completely suppresses the influence of laser frequency noise, which can be a limiting factor in AMs operating away from optical resonance \cite{Petrenko2021}.

The proposed scheme is relatively simple to implement, which will facilitate a high degree of sensor miniaturization. In particular, there is no need for optical filters or other auxiliary components to block the pump beam, as in some single-beam pump-probe schemes \cite{Johnson2010,Zhao2023}. The technique proposed does not rely on the SERF regime, so that the cell temperature can be relatively low ($\lesssim\,$$85^\circ$C). This allows for a reduction in both the overall power consumption of the AM and the heat release of the sensor head, which is advantageous for biomedical applications.

\section{Theoretical model}\label{sec:2}

\begin{figure}[!t]
\includegraphics[width=1\linewidth]{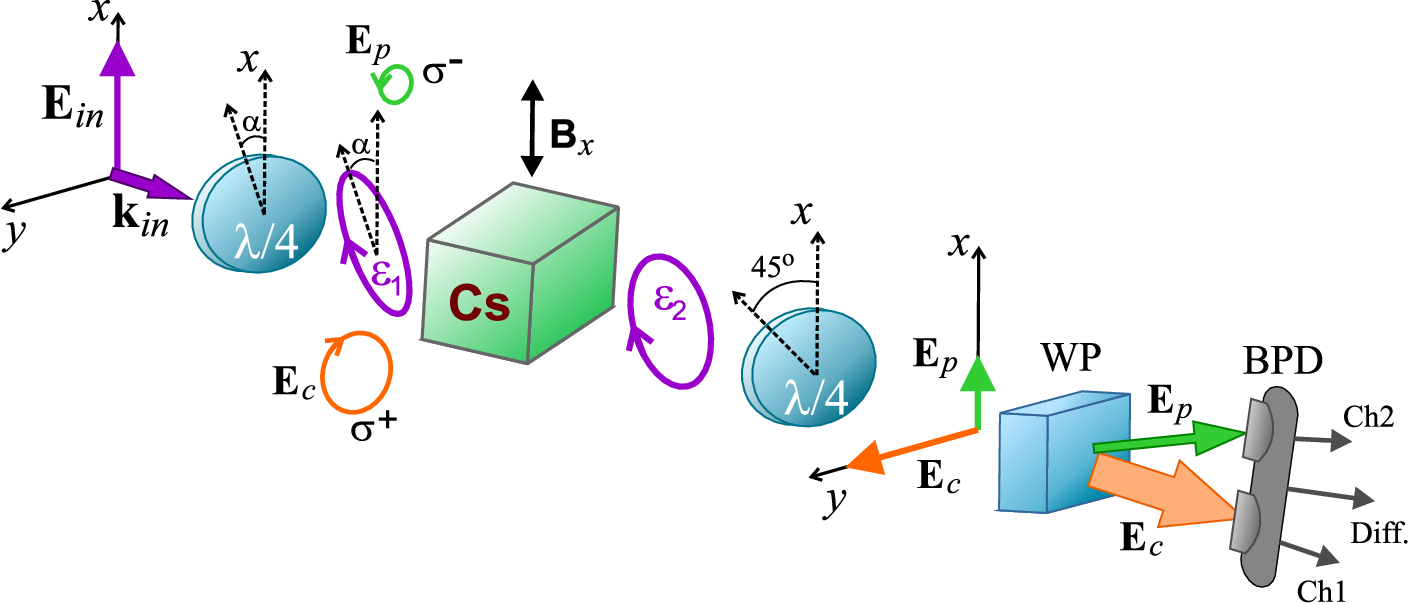}
\caption{\label{fig:1}Transformation of the light wave polarization upon passing through the sensor head: $\lambda/4$, quarter-wave plate; Cs, cesium vapor cell; WP, Wollaston prism; BPD, balanced photodetector; Ch1, Ch2, Diff, photodetector outputs corresponding to channel~1, channel~2, and differential (balanced) channel, respectively. ${\bf E}_{\it in}$ is the linearly polarized input wave with the wave vector ${\bf k}_{\it in}$.}
\end{figure}

Let us consider the resonant interaction of alkali atoms with an elliptically polarized light wave propagating along the quantization axis $z$ (see Fig.~\ref{fig:1}):

\begin{equation}\label{lightfield}
{\bf E}(z,t) = E_0\,{\bf e}\,e^{-i(\omega t - k z)} + c.c.
\end{equation}

\noindent Here, $E_0$ is the real amplitude of the light wave, related to the intensity as $I = (c/2\pi) E_0^2$, while $\omega$ and $k$ denote the optical frequency and the magnitude of the wave vector, respectively, and ``{\it c.c.}'' stands for the complex conjugate term. The complex unit polarization vector ${\bf e}$ in the spherical basis is given by:

\begin{equation}\label{polarization}
{\bf e} = -\sin(\epsilon-\pi/4)\,e^{i\varphi}\,{\bf e}_{-1} -\cos(\epsilon-\pi/4)\,e^{-i\varphi}\,{\bf e}_{+1}\,,
\end{equation}

\noindent where ${\bf e}_{\pm1}$ are the unit vectors of the spherical coordinate system, related to the Cartesian one as follows \cite{Varshalovich1988}: ${\bf e}_{\pm1} = \mp({\bf e}_x \pm i {\bf e}_y)/\sqrt{2}$ with $\epsilon$ being the polarization ellipticity parameter, $-\pi/4 \leq \epsilon \leq \pi/4$. In this notation, $\epsilon = 0$ corresponds to linearly polarized light, while $\epsilon = \pm \pi/4$ corresponds to right-handed ($+$) and left-handed ($-$) circular polarization. The angle $\varphi$ denotes the orientation of the major axis of the polarization ellipse with respect to the $x$ axis.

At the entrance to the sensor head of the AM, the light is linearly polarized (see Fig.~\ref{fig:1}). The conversion of linear to elliptical polarization is achieved by the first quarter-wave plate, whose fast axis is rotated by an angle $\alpha$ around the $z$ axis, such that after the plate $\epsilon$$\,=\,$$\varphi$$\,=\,$$\alpha$. For our analysis, it is convenient to represent the wave (\ref{lightfield}) as a superposition of two components: the pump wave ${\bf E}_c$, assumed to be right-circularly polarized and denoted as $\sigma^+$ (orange circle in Fig.~\ref{fig:1}), and the probe wave ${\bf E}_p$, which is left-circularly polarized ($\sigma^-$, green circle in Fig.~\ref{fig:1}). The amplitudes of these waves are given by $E_c = E_0\,{\rm cos}(\epsilon-\pi/4)$ and $E_p = E_0\,{\rm sin}(\epsilon-\pi/4)$, with the corresponding intensities $I_{c,p} = cE_{c,p}^2/2\pi$.

The pump and probe waves are mutually coherent, being components of a single elliptically polarized wave. In this case, their interaction with atoms can induce Zeeman coherences in the atomic ground state, which are superpositions of states with the same total angular momentum $F_{g}$ but different magnetic quantum numbers $m_g$. However, in vapor cells with relatively high buffer-gas pressure, where the hyperfine excited-state manifold is not spectrally resolved, the effects of such Zeeman coherences are strongly suppressed. Such buffer-gas pressures ($\gtrsim\,$$100$~Torr) are employed in the vast majority of miniaturized magnetic sensors, as well as in our experiments. Thus, light-induced Zeeman coherences can be neglected, allowing the two field components, ${\bf E}_c$ and ${\bf E}_p$, to be considered as independent.

Since the photons of ${\bf E}_c$ and ${\bf E}_p$ waves possess spins oriented in opposite directions (parallel and antiparallel to the $z$ axis, respectively), a form of ``competition'' arises between these waves in orienting the atomic spins during optical pumping. In atomic magnetometry, a qualitative description of optical spin orientation (ground-state polarization) and its subsequent precession in an external magnetic field is often provided by an intuitive and simplified formalism based on the Bloch equation for the magnetization vector ${\bf M}$, which represents the dipole magnetic moment per unit volume of the medium \cite{Dehmelt1957,Slocum1973}. In the case of two waves, ${\bf E}_c$ and ${\bf E}_p$, and a magnetic field ${\bf B}$ directed along the $x$ axis, this equation can be written as

\begin{eqnarray}\label{BlochEq}
    \frac{d{\bf M}}{dt} =&&\, \gamma \bigl[{\bf M}\times{\bf B}_x\bigr] + R_c\,\bigl({\bf M}_0-{\bf M}\bigr) \nonumber\\
    &&\qquad\qquad  - R_p\,\bigl({\bf M}_0+{\bf M}\bigr) - \Gamma\, {\bf M}\,.
\end{eqnarray}

\noindent Here, $\gamma$$\,\approx\,$$2\pi$$\times$$3.5$~Hz/nT is the gyromagnetic ratio for the cesium atom, while $R_c$ and $R_p$ denote the optical pumping rates of the atomic ground state induced by the waves ${\bf E}_c$ and ${\bf E}_p$, respectively ($R_{c,p}$$\,\propto\,$$I_{c,p}$). For the sake of simplicity, we assume that the transverse and longitudinal relaxation rates are equal, and thus a single relaxation constant $\Gamma$ is introduced to describe these processes. In Eq.~(\ref{BlochEq}), it is further assumed that in the steady-state regime ($\dot{{\bf M}}$$\,=\,$$0$), in the absence of the magnetic field, relaxation, and probe wave, the wave ${\bf E}_c$ induces a magnetic moment in the medium of the form ${\bf M}_0$$\,=\,$$M_0$$\,{\bf e}_z$. Under the same conditions, the wave ${\bf E}_p$ induces a moment $-{\bf M}_0$ (where ${\bf e}_z$ is the Cartesian unit vector along the $z$ axis).

In the experiments, ${\bf B}_x$ is scanned at a certain frequency to observe a MOR. We assume this frequency to be sufficiently low such that the adiabatic regime applies, in which the magnetic field in Eq.~(\ref{BlochEq}) can be treated as a time-independent parameter. A more rigorous treatment of the problem has been presented in numerous works, but it is unnecessary for our qualitative analysis (e.g., see \cite{Slocum1973}). Thus, in our qualitative theoretical model, we consider the steady-state regime, i.e. $\dot{{\bf M}}$$\,=\,$$0$.

The mathematical details of solving Eq.~(\ref{BlochEq}) are provided in Supplementary Material. In particular, to determine the intensities of the waves ${\bf E}_c$ and ${\bf E}_p$ after the cell, it is sufficient to consider only the $z$-component of the vector ${\bf M}$, which is described by a Lorentzian profile:

\begin{equation}
    M_z = \frac{\Delta(R_c - R_p)}{\Delta^2 + \Omega^2}M_0\,,
    \label{SolutionMz}
\end{equation}

\noindent where $\Omega$$\,=\,$$\gamma B_x$ is the Larmor frequency, and $\Delta$ is the half-width at half-maximum (HWHM) of the resonance, including power broadening:

\begin{equation}\label{Width}
    \Delta = \Gamma + R_c + R_p \,.
\end{equation}

\noindent Spin-exchange relaxation, which is a part of $\Gamma$, can be reduced due to the high degree of spin orientation of atoms under the action of the pumping wave. Therefore, $\Gamma$ also depends on the light intensity. This effect is known as light narrowing of the resonance \cite{Savukov2005}. From Eq.~(\ref{SolutionMz}), it follows directly that if the intensities of the two waves are equal, i.e., $R_c$$\,=\,$$R_p$, no macroscopic magnetic moment is induced in the medium because of the aforementioned competition between the two waves.

During propagation through the medium, the intensities of the waves ${\bf E}_c$ and ${\bf E}_p$  obey the Beer-Bouguer-Lambert law. For our qualitative analysis, we consider the approximation of an optically thin medium, in which $M_z$ does not significantly depend on $z$. Under this assumption, we get the following expressions for the intensities of the light waves after the vapor cell (see Supplementary Material):

\begin{eqnarray}
    &&I_c(L) \approx I_c(0) \Biggl[ 1 - \varkappa L - \varkappa L \frac{\Delta(R_c - R_p)}{\Delta^2 + \Omega^2} \Biggr]\,,\label{IntensC}\\
    &&I_p(L) \approx I_p(0) \Biggl[ 1 - \varkappa L + \varkappa L \frac{\Delta(R_c - R_p)}{\Delta^2 + \Omega^2} \Biggr]\,,
    \label{IntensP}
\end{eqnarray}

\noindent where $I_{c,p}(0)$ and $I_{c,p}(L)$ are the wave intensities at the entrance to and exit from the vapor cell, respectively, $L$ is the cell length, and $\varkappa$ is the light absorption coefficient for interaction with unpolarized atoms, i.e., under the condition $M_z$$\ll$$M_0$. The sign preceding the fractions in (\ref{IntensC}) and (\ref{IntensP}) depends on which hyperfine ground-state manifold is being excited by the resonant light (see Supplementary Material). In our experiments, the light beam excites Cs atoms from the lowest ground-state hyperfine level with $F_g$$\,=\,$$3$.

\begin{figure}[!t]
\includegraphics[width=1\linewidth]{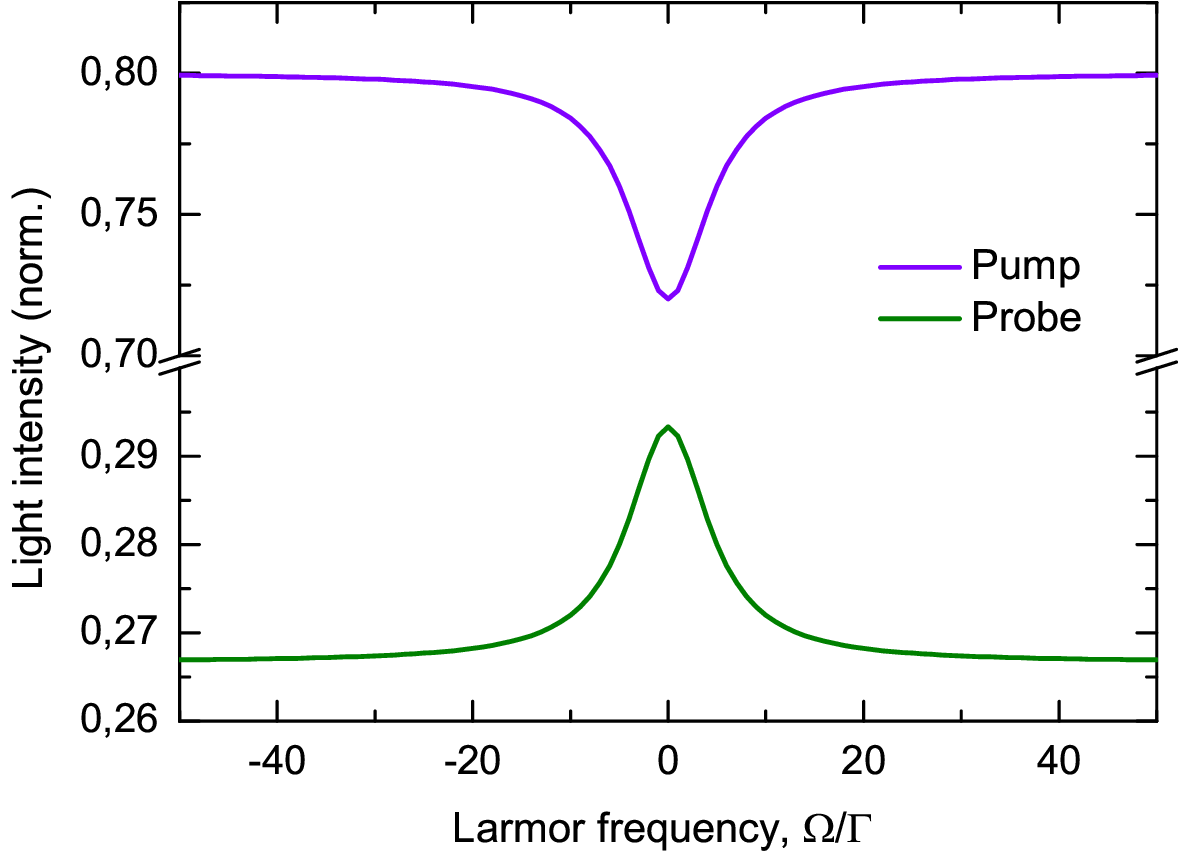}
\caption{\label{fig:2}The transmitted intensity of the pump wave (upper curve) and the probe wave (lower curve) after passing through the vapor cell as a function of the transverse magnetic field (Larmor frequency in units of $\Gamma$). Parameters used for the calculation as follows: $I_p(0)$$\,=\,$$I_c(0)/3$, $R_c$$\,= \,$$3R_p$$\,=\,$$3\Gamma$, $\varkappa L$$\,=\,$$0.2$. The curves are normalized to $I_c(0)$.}
\end{figure}

We have not yet imposed any restrictions on the ratio of the pumping and probe wave intensities. However, in experiments $I_c$ exceeds $I_p$ by several times (see Sec.~\ref{sec:3}). Fig.~\ref{fig:2} shows the dependencies (\ref{IntensC}), (\ref{IntensP}) for such a case, where the optical pumping rate $R_c$ is three times larger than $R_p$. As can be seen from the figure, the resonances exhibit opposite signs: in the transmission of the pumping wave, an electromagnetically induced absorption (EIA) resonance is observed, whereas in the transmission of the probe one an electromagnetically induced transparency (EIT) resonance occurs. The behavior of these resonances is explained by the circular magnetically sensitive dichroism of the medium \cite{Brazhnikov2022}. These resonances share many similarities with previously studied effects in the field of counter-propagating circularly polarized waves \cite{Brazhnikov2021} and in the field of linearly polarized waves \cite{Brazhnikov2019,Brazhnikov2020,Makarov2023}, where linear dichroism manifested in a similar way. Qualitative arguments explaining the signs of the resonances for different types of optical transitions are presented in Supplementary Material.

In the experiments, in order to observe the transmission signals shown in Fig.~\ref{fig:2}, it is necessary to spatially separate the waves ${\bf E}_c$ and ${\bf E}_p$ after passing through the cell. In the proposed scheme, this is achieved by means of an additional $\lambda/4$ plate placed after the cell, which converts the circular polarizations of the pump and probe waves into mutually orthogonal linear polarizations. A Wollaston prism then separates these waves into two beams and directs them to the balanced photodetector (BPD). At its output channel 1 (``Ch1'' in Fig.~\ref{fig:1}), the transmission signal of the ${\bf E}_c$ wave can be observed, while channel 2 (``Ch2'') provides the signal from the ${\bf E}_p$ wave (the signals are provided in Supplementary Material).


The opposite signs of the observed EIT and EIA resonances allow one signal to be subtracted from the other, thereby enhancing the amplitude of the MOR and significantly suppressing the light intensity noise. To obtain the differential signal, we use the balanced output of the photodetector (``Diff.''). The intensity noise of the laser radiation incident on the photodetector is proportional to the intensity itself. Therefore, subtraction of the noise in the balanced channel becomes more effective when the two observed signals in Ch1 and Ch2 are closer in magnitude. In particular, in Fig.~\ref{fig:2}, the background intensity, i.e., the ``pedestal'' of the resonance, is approximately $0.27$ for the probe wave and $0.8$ for the pump wave. Before subtracting these signals, the intensities of the two beams must be equalized. This can be accomplished in various ways, for example, through subsequent signal processing by an electronic system, as demonstrated in \cite{Oelsner2019,Chen2022,Ma2024}. However, no significant phase delay should occur between the subtracted signals within the frequency band of interest (from dc up to a few kHz). In our experiments, this task was solved in the simplest way by inserting an additional $\lambda/2$ plate before the Wollaston prism (not shown in Fig.~\ref{fig:1}). This plate enables redistribution of the wave intensities between the two BPD channels, ensuring equal background intensities in each of them.

It should be noted that the theoretical analysis presented in this section is intended only as a qualitative explanation of the signals observed in the experiments. A more rigorous treatment, however, should be based on the density matrix approach and should take into account not only dichroism effects but also birefringence effects, since both phenomena contribute to the signals detected by the polarimeter \cite{Shah2009,Nienhuis1998}. Moreover, quenching of atomic excited-state populations due to collisions between Cs and N$_2$ may increase an impact of the birefringence effect due to efficient orientation of the atomic spins on the adjacent hyperfine ground-state level \cite{Popov2018}. Nevertheless, all these features deserve a separate comprehensive theoretical study and lie beyond the scope of the present work.

\begin{figure*}[!t]
\includegraphics[width=1\linewidth]{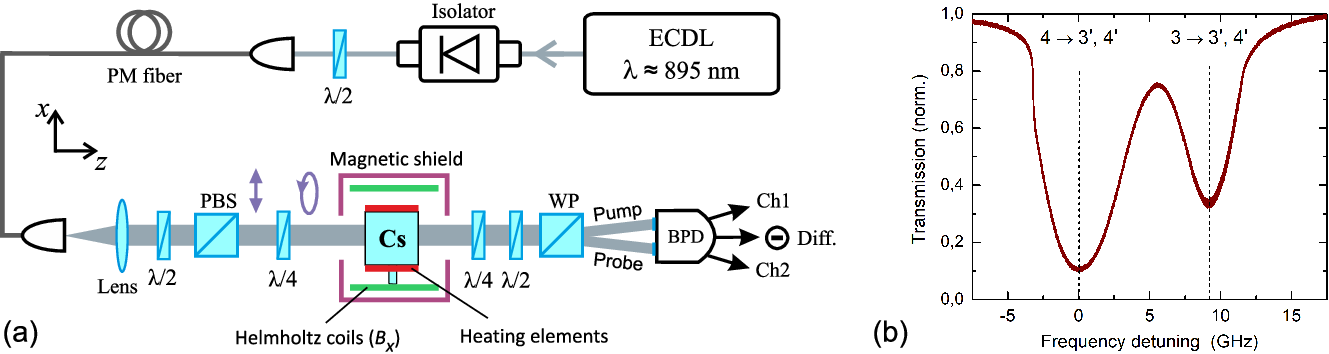}
\caption{\label{fig:3}(a) Schematic of the experimental apparatus: ECDL, external-cavity diode laser; $\lambda/2$, $\lambda/4$, half- and quarter-wave plates, respectively; PM fiber, polarization-maintaining fiber; PBS, polarizing beam splitter; Cs, cesium vapor cell containing buffer gas; WP, Wollaston prism; BPD, balanced photodetector. (b) Measured absorption profile of the vapor cell at $T$$\,\approx\,$$80\,^\circ$C and $P$$\,\approx\,$$1$~mW.}
\end{figure*}

\section{Experiments}\label{sec:3}

Figure \ref{fig:3}(a) shows a schematic of the experimental setup. MORs were studied using a Littrow-configuration external-cavity diode laser (ECDL) with a spectral linewidth below $1$~MHz \cite{VitaWave}. Continuous wavelength tuning was achieved by means of a piezoceramic-mounted diffraction grating. The wavelength was monitored using a WS7 wavemeter by ``Angstrom'' Ltd.

The laser beam was passed through a Faraday optical isolator to suppress parasitic back reflections and coupled into a polarization-maintaining optical fiber. Using the fiber collimator and an additional lens, the beam diameter was expanded to approximately $3$~mm (FWHM). The radiation then passed through a half-wave plate ($\lambda/2$) and a polarizing beam splitter (PBS), which together allowed smooth adjustment of the optical power. A quarter-wave plate ($\lambda/4$) placed in front of the vapor cell was used to control the ellipticity of the light polarization. The beam was then directed into a cubic $5$$\times$$5$$\times$$5$~mm$^3$ glass cell containing $^{133}$Cs vapor. At the relatively high buffer-gas pressure of about 200 Torr (Ar:N$_2$$\,\approx\,$$80$$:$$120$), the hyperfine splitting of the excited state of cesium was not spectrally resolved [Fig. \ref{fig:3}(b)]. Using the wavemeter, the laser frequency was tuned to the center of the absorption profile corresponding to the $F_g$$\,=\,$$3$$\,\to\,$$F_e$$\,=\,$$3',$$4'$ optical transitions.
It should be noted that the method proposed can be applied to various types of optical transitions in alkali-metal atoms, both ``bright'' transitions where $F_g$$\,=\,$$F$$\,\to\,$$F_e$$\,=\,$$F$$+$$1$, as well as ``dark'' transitions where $F_g$$\,=\,$$F$$\,\to\,$$F_e$$\,=\,$$F$$-$$1$ or $F_g$$\,=\,$$F$$\,\to\,$$F_e$$\,=\,$$F$. In our experiments, the measured sensitivity for excitations of atoms with $F_g$$\,=\,$$3$ and $F_g$$\,=\,$$4$ was found to be comparable. Therefore, we present results only for $F_g$$\,=\,$$3$.

For the observation of MORs, a transverse magnetic field (${\bf B}$$\perp$${\bf k}$) was slowly ($\sim$$1$~Hz) scanned around zero. The field was produced by a pair of Helmholtz coils supplied by a DG4162 generator (``Rigol Technologies''). The cell together with the coils was placed inside a three-layer magnetic shield, reducing the laboratory magnetic field at the cell position to below $50$~nT. Additional Helmholtz coils were utilized to further compensate a residual magnetic field. Heating of the atomic cell was provided by a $100$~kHz current applied to thin bifilar resistive elements. The heating system produced no observable effect on the width and amplitude of the resonances.

\begin{figure*}[!t]
\includegraphics[width=\linewidth]{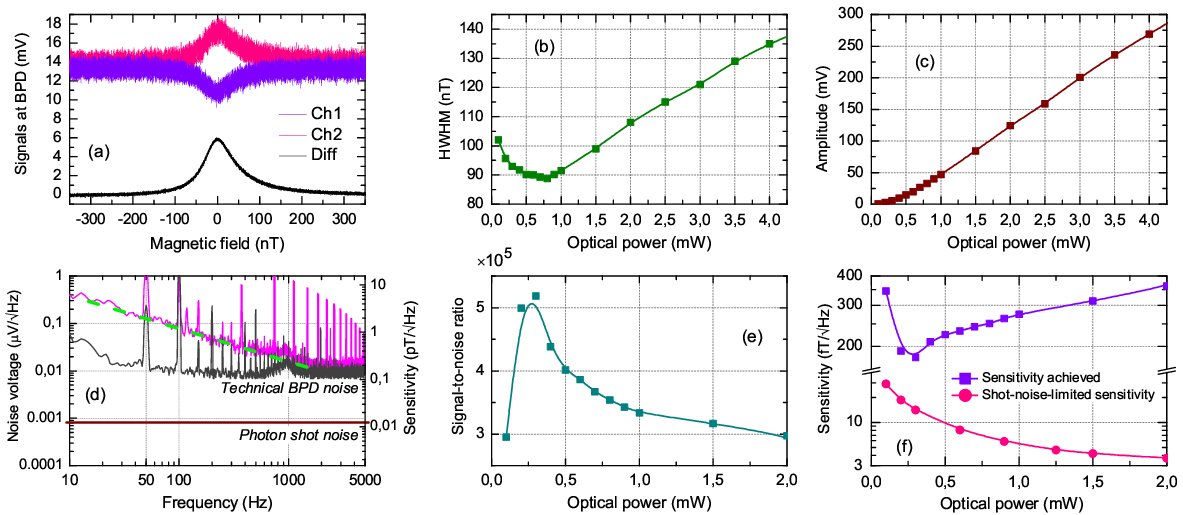}
\caption{\label{fig:4}(a) MOR signals recorded from the three BPD channels at $P$$\,\approx\,$$300$~$\mu$W (the photodetector bandwidth is $\approx\,$$1$~MHz). (b), (c) The half-width at half maximum and amplitude of the resonance as functions of the optical power at the input of the vapor cell. (d) Signal noise from the balanced BPD channel (magenta) measured on the resonance slope at $P$$\,\approx\,$$300$~$\mu$W. The contribution of the technical noise of the photodetector is shown as black data, while the photon-shot-noise limit is shown as a horizontal solid line. (e), (f) Signal-to-noise ratio and sensor sensitivity as functions of optical power, calculated according to Eq. (\ref{Sensitivity}). In all measurements $T$$\,\approx\,$$85$$\,^\circ$C and $\epsilon$$\,\approx\,$$10^\circ$. Solid lines in panels (b), (c), (e), and (f) are drawn as a guide to the eye.}
\end{figure*}

Figure \ref{fig:4}(a) shows MORs recorded from the three channels of the BPD (PDB210A by ``Thorlabs''). The resonance noise observed on the oscilloscope was mainly limited by the bandwidth of the balanced output of the photodetector, which is approximately $1$~MHz. As can be seen, the balanced channel exhibits a significantly lower noise compared to the individual channels, which results from efficient subtraction of the laser intensity noise. Figures \ref{fig:4}(b,c) present the characteristics of the MORs in the balanced channel. In particular, Fig. \ref{fig:4}(b) demonstrates the effect of light-induced narrowing with increasing optical power, observed up to $800$~$\mu$W, where HWHM equals $\approx\,$$90$~nT. This value is, as expected, higher than that typically observed in miniature SERF sensors. For instance, the commercial QuSpin QZFM sensor exhibits a resonance half-width of about 15 nT \cite{Osborne2018}. However, as will be shown below, in our case high measurement sensitivity can still be achieved owing to the large signal-to-noise ratio. At optical powers above $800$~$\mu$W, Fig. \ref{fig:4}(b) shows power broadening of the resonance, scaling linearly with power, in qualitative agreement with Eq. (\ref{Width}). As illustrated in Fig. \ref{fig:4}(c), the MOR amplitude also increases almost linearly with optical power, consistent with the theoretical predictions given by Eqs. (\ref{IntensC}) and (\ref{IntensP}).

Figure \ref{fig:4}(d) shows the noise spectrum of the balanced BPD channel recorded at a magnetic field of $\approx$$\,90$~nT, corresponding to the resonance slope. The measurement was performed with an SR1 spectrum analyzer (``Stanford Research Systems''). The distinct high peaks are clearly seen in the spectrum, coming from the ac line $50$~Hz and its frequency harmonics. The spectrum also reveals technical noise that follows the typical $\sim\,$$1/f^\eta$ law (green dashed line). Its cutoff frequency is about $1.5$~kHz. Possible sources of the noise measured below $1.5$~kHz include heating system of the cell, electric current noise in the Helmholtz coils, and insufficient shielding of the cell from low-frequency magnetic field noise \cite{Fedosov2025}. These types of noise cannot be reduced by means of observation of a differential signal. It will be shown below that the primary contribution to the measured noise above $\approx\,$$1$~kHz is made by the photodetector.

It is important to note that all mentioned technical noises can be substantially reduced during the development of a magnetic field sensor. For instance, magnetic noise in the cell can be reduced by using additional magnetic shields, including a ferrite layer \cite{Kornack2007,Lu2020}, or by employing a gradiometric sensor configuration \cite{Petrenko2022}, which allows to reduce a common-mode detection noise in two vapor cells by more than an order of magnitude. The current noise in the Helmholtz coils can also be significantly reduced by employing a specialized low-noise generator instead of a general-purpose one.

Fig. \ref{fig:4}(d) shows the photodetector intrinsic noise. This trace was obtained using an incandescent lamp providing the same photodetector output voltage ($\approx\,$$14$~mV) as during MOR recording at $300$~$\mu$W [see Fig. \ref{fig:4}(a)]. The noise from the incandescent lamp radiation in the low-frequency range is expected to represent solely white noise, independent of frequency. The obtained data indicate that the noise spectrum also contains a component at $50$~Hz and many of its harmonics. Moreover, above $1$~kHz the noise observed in the lamp experiment closely follows the noise measured on the slope of the magneto-optical resonance. From these measurements it can be concluded that, under the current experimental conditions, the noise of the magneto-optical resonances above $1$~kHz is determined primarily by the photodetector itself. Note that this level is not determined by the photocurrent shot noise, which is the minimum possible noise of a photodetector. In our case, the observed photodetector noise seems to be caused by the thermal (Johnson-Nyquist) noise of internal components. Fig. \ref{fig:4}(d) illustrates the minimum achievable noise level in our setup using this photodetector, which equals to $\approx\,$$10$~nV/$\surd$Hz.

\begin{figure*}[!t]
\includegraphics[width=0.8\linewidth]{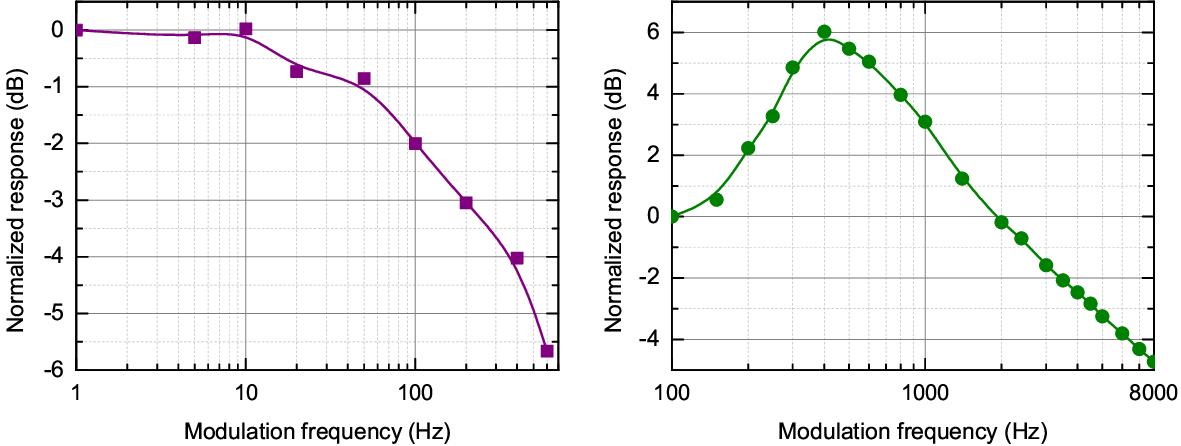}
\caption{\label{fig:5}(a) Resonance slope as a function of the modulation frequency at a fixed scan amplitude. (b) Resonance slope as a function of the modulation frequency at a fixed modulation index, i.e., the ratio of amplitude to modulation frequency. $T$$\,\approx\,$$85\,^\circ$C, $P$$\,\approx\,$$500$~$\mu$W, $\epsilon$$\,\approx\,$$10^\circ$. Solid lines are drawn as a guide to the eye.}
\end{figure*}

Fig. \ref{fig:4}(e) shows the dependence of the $1$-Hz bandwidth signal-to-noise ratio (SNR) on optical power. To estimate the SNR, we measured the signal noise at the frequency of $2$~kHz (the explanations will be provided below). The maximum SNR reaches the value of $\approx\,$$5$$\,\times\,$$10^5$ at $P$$\,\approx\,$$300$~$\mu$W. The sensitivity of a magnetic sensor, defined as the minimum detectable magnetic field variation in a $1$~Hz bandwidth, was estimated using the common expression \cite{Martinez2010}:

\begin{equation}\label{Sensitivity}
\delta B \approx \frac{\Delta}{{\rm SNR}}\,.
\end{equation}

\noindent The dependence $\delta B(P)$ is shown in Fig. \ref{fig:4}(f). Under the current experimental conditions, the optimal sensitivity of the sensor is about $180$~fT/$\surd$Hz. This sub-picotesla sensitivity of the sensor is more than sufficient, for example, for performing MCG, where it is necessary to record magnetic fields ranging from approximately $10$ to $100$~pT \cite{Xiao2023,Chen2025}.

However, the ultimate performance of a magnetic sensor is commonly evaluated in terms of the photon shot noise or the projection shot noise of the atoms \cite{Fabricant2023,Vershovskii2020,Budker2023}. The latter is proportional to the square root of the number of atoms contributing to the resonance. This noise is usually one of the main sensitivity limiting factors in extremely small vapor cells, whereas for cells with a volume greater than $0.1$~cm$^3$, as in our experiments, the photon shot noise becomes a dominant limiting factor. The SNR in the photon-shot-noise limit approximately equals to the square root of the photon flux, i.e., the number of photons per second \cite{Fabricant2023,Budker2023}. Thus, for the resonance shown in Fig. \ref{fig:4}(a), SNR approaches $\approx\,$$7$$\times$$10^6$ (the corresponding noise level is shown in Fig. \ref{fig:4}(d) as a horizontal solid line). With $\Delta$$\,\approx\,$$90$~nT, this yields an estimate of the ultimate sensor sensitivity of $\approx\,$$15$~fT/$\surd$Hz. However, with increasing the light power, the ultimate sensor sensitivity can be much better. In Fig. \ref{fig:4}(f), it reaches a floor of $\approx\,$$3$~fT/$\surd$Hz at $P$$\,\approx\,$$3$~mW. It is known that the ultimate sensitivity of a sensor can be very close to such a floor even in a low-frequency band \cite{Petrenko2022,Klassen2024}. Moreover, the noise can be below this limit when the light squeezing effect is engaged \cite{Horrom2012,Li2022}. Therefore, after reduction of the technical noise in our experimental apparatus, we expect to achieve a sensitivity of about $5$~fT/$\surd$Hz (a contribution from the Johnson-Nyquist noise from thermal electron motion in the magnetic shield and Helmholtz coils is estimated to be of the order of a few fT/$\surd$Hz \cite{Lee2008}). Such a sensitivity is nowadays demonstrated by the world's most advanced miniaturized GSHE-based magnetic sensors used for medical diagnostics and operating in the SERF regime \cite{Osborne2018,Alem2023}.

To know bandwidth of the sensor, we have measured the resonance modulation response [Fig. \ref{fig:5}(a)]. In this experiment, the magnetic field $B_x$ was set to a fixed offset corresponding to the resonance slope ($\approx\,$$90$~nT). The field was then modulated with a fixed deviation of $\Delta B$$\,\approx\,$$15$~nT at various frequencies $f_m$. The amplitude of the corresponding modulated signal at the balanced BPD channel was recorded. As $f_m$ increased, the slope of the resonance gradually decreased, resulting in a reduction of the modulation amplitude. The modulation frequency $f_m$ at which this amplitude drops by a factor of $2$ ($-3$~dB) relative to the low-frequency region ($\sim\,$$1$~Hz) defines the sensor bandwidth. As seen from Fig. \ref{fig:5}(a), in our case, the bandwidth is $\approx$$\,200$~Hz. It should be noted that these measurements are equivalent to evaluating the slope of the error signal \cite{Shah2009}, which is formed in the magnetometer by means of a lock-in amplifier using the standard modulation-demodulation technique. Near $B_x$$\,\approx\,$$0$, the error signal appears as a straight line with a slope, which is proportional to the ratio of the resonance amplitude to its width. We can expect that the bandwidth will be increased further, if the sensor operates in a closed-loop regime \cite{Alem2023}.

The sensor bandwidth does not strictly determine the maximum modulation frequency that can be used in a magnetometer to form the error signal. Indeed, the slope of the error signal is determined by the modulation index of the harmonic voltage applied to the Helmholtz coils, defined as the ratio $\mu$$\,=\,$$\Delta B$$/$$f_m$ \cite{Shah2009,Cohen1970}. Therefore, increasing the modulation frequency reduces both $\mu$ and the slope. However, to restore the original value of $\mu$ and, consequently, the slope of the error signal, it is sufficient to increase $\Delta B$ proportionally with $f_m$. Thus, it is possible to generate an error signal with a high slope using a modulation frequency significantly higher than the sensor bandwidth. For example, in Ref. \cite{Shah2009}, a modulation frequency above $2$~kHz was used for a sensor with a bandwidth of only about $150$~Hz. Clearly, maintaining a high error-signal slope in this manner is feasible only until $\Delta B$ becomes comparable to the half width of the resonance. Further simultaneous increase of $\Delta B$ and $f_m$ inevitably lead to a reduction of the error-signal slope and a corresponding decrease in sensor sensitivity. As shown in Fig. \ref{fig:5}(b), in our setup the modulation frequency can be set to $2$~kHz without any degradation of the error signal relative to the low-frequency region ($<\,$$100$~Hz). This is the reason why the noise voltage in our experiments was measured at the frequency of $2$~kHz to estimate the sensor sensitivity.

As seen in Fig. \ref{fig:5}(b), with increasing the modulation frequency and scanning amplitude, the resonance slope demonstrates an increase up to $f_m$$\,\approx\,$$400$~Hz. This effect is associated with at least two reasons. First, in the low-frequency region, the resonance slope degrades only slightly with increasing scan frequency. Therefore, increasing the scanning amplitude leads to a corresponding increase in the amplitude of the observed harmonic signal. Second, there can be a specific distortion of the Lorentzian lineshape of the MOR, which leads to an enhanced resonance slope. Similar distortions of a Hanle resonance lineshape have been observed in previous studies when the scanning frequency was increased while the scanning amplitude was not small compared to the resonance width \cite{Li2023,Zhao2024}.

\section{Conclusion}


In this work, a novel scheme for observation of zero-field level-crossing resonances was investigated. The experimental results demonstrate the potential of the scheme for the development of a miniature high-sensitivity magnetic sensor operating in ultra-low magnetic fields. A key feature is that high sensitivity of measurements can be achieved without employing the SERF regime, i.e., at significantly lower temperature of alkali-metal vapor. This distinguishes the proposed scheme from the vast majority of state-of-the-art commercial miniature GSHE-based sensors. In our scheme, the observations of the magneto-optical resonances were performed using a polarimeter composed of such optical elements as quarter- and half-wave plates, as well as a Wollaston prism. However, for the purpose of further ultimate miniaturization of the sensor head to a volume of $\sim\,$$1$~cm$^3$, these elements can be replaced by a single nanophotonic device, such as a spin selector \cite{Sebbag2021}.

Under the current experimental conditions, the sensitivity of magnetic field measurements in the proposed scheme is estimated to be approximately $180$~fT/$\surd$Hz. Such a sensor can be applied for performing MCG. At the moment, the sensitivity is mainly limited by technical noise which can be overcome when developing a sensor. Thus, the ultimate sensitivity of the sensor is estimated to be approximately $5$~fT/$\surd$Hz. In addition to the drastic reduction of the cell temperature and hence the heat dissipation of the sensor head, we have measured an extended bandwidth equaled to $200$~Hz. In the future, this value can be doubled by using the closed-loop regime of operation \cite{Alem2023}. Since the linewidth of the resonance in our scheme is several times larger than that in SERF magnetic sensors, the dynamic range is correspondingly increased, and the requirements for magnetic shielding are significantly relaxed (e.g., see \cite{Zhang2025}). All the listed features of the proposed technique make it attractive for applications in medicine and biology, competing with state-of-the-art GSHE-based sensors operating in the SERF regime.

\begin{acknowledgments}

We thank V.I.~Vishnyakov for the help with electronic systems used in the experiments. The work was supported by the Russian Science Foundation (Grant \# 23-12-00195).

\end{acknowledgments}


\appendix*

\section{Theoretical analysis}

To study the ground-state Hanle effect (GSHE) in an elliptically polarized light wave, we use a common simplified approach based on the Bloch equation in the form (\ref{BlochEq}). It is convenient to rewrite this equation in the following way \cite{Slocum1973}:

\begin{eqnarray}
    &&\frac{dM_x}{dt} = -\bigl(\Gamma + R_c + R_p\bigr)M_x\,,\label{BlochSystem1}\\
    &&\frac{dM_+}{dt} = i\Omega M_+ - \bigl(\Gamma + R_c + R_p\bigr)M_+ \nonumber\\
    &&\qquad\qquad\qquad\qquad\qquad  + \bigl(R_c - R_p\bigr)M_0\,,
    \label{BlochSystem2}
\end{eqnarray}

\noindent where $M_+$$\,=\,$$M_z$$\,+\,$$iM_y$. Thus, two independent equations arise: the first one for the longitudinal component (${\bf M}_x$$||$${\bf B}_x$), and the second involving only the transverse components of the vector ${\bf M}$.

\begin{figure*}[!t]
\includegraphics[width=0.7\linewidth]{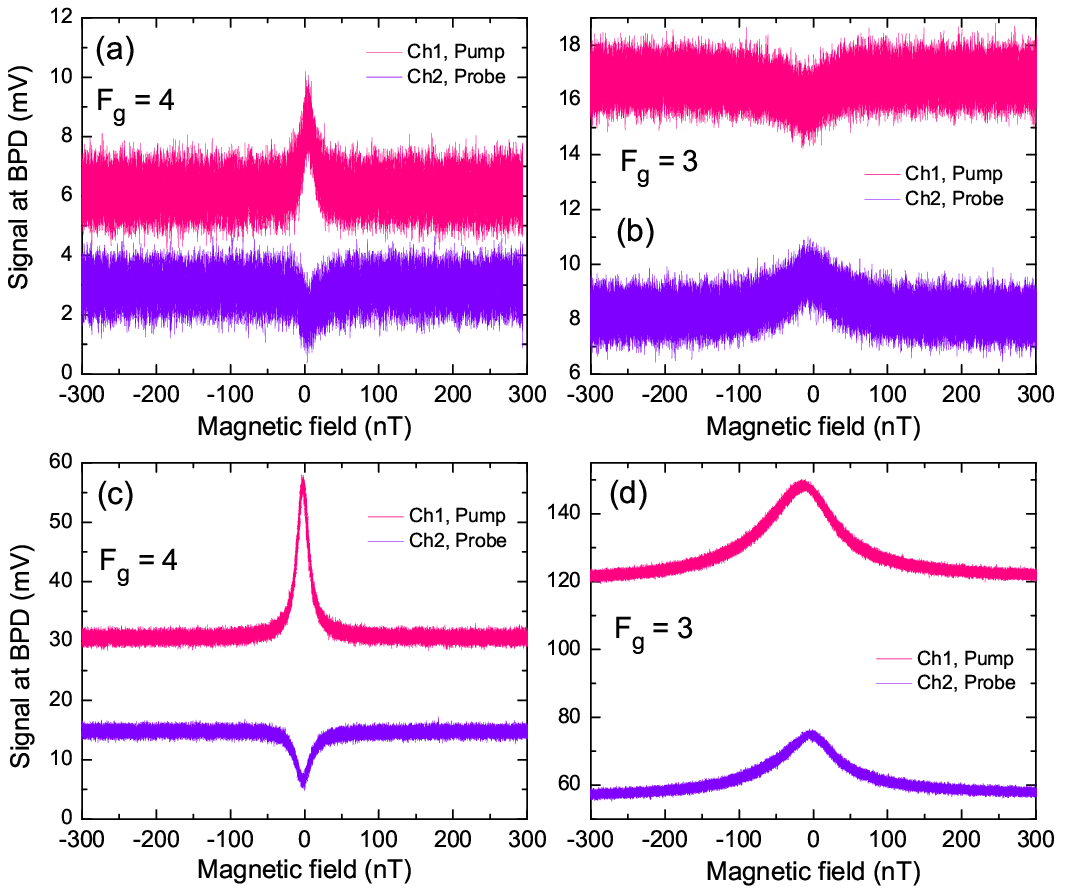}
\caption{\label{fig:6}Magneto-optical resonances observed in the cesium vapor cell: (a,c) $F_g$$\,=\,$$4$ level or (b,d) $F_g = 3$ level is excited. (a,b) $P$$\,\approx\,$$400$~$\mu$W, (c,d) $P$$\,\approx\,$$1000$~$\mu$W. Other parameters: $T$$\,\approx\,$$85^\circ$C, $\epsilon$$\,\approx\,$$10^\circ$. The upper curve represents MOR in channel 1 (pump wave transmission), while the lower curve represents MOR in channel 2 (probe wave transmission) of the balanced photodetector.}
\end{figure*}

We consider ${\bf B}_x$ to be slowly scanned around zero value, so that it can be treated as one more parameter in (\ref{BlochEq}). Moreover, to explain qualitatively experimental curves, it is enough to consider a steady-state regime when $\dot{M}_x$$\,=\,$$\dot{M}_+$$\,=\,$$0$. Then, from equations (\ref{BlochSystem1}) and (\ref{BlochSystem2}), the solution is obtained immediately as follows:

\begin{eqnarray}
    &&M_x = 0\,,\\
    &&M_+ = \frac{R_c - R_p}{\Gamma + R_c + R_p - i\Omega}M_0\,.
    \label{Solution2}
\end{eqnarray}

\noindent Using the relation $M_z$$\,=\,$$\text{Re}$$\bigl\{$$M_+$$\bigr\}$ and introducing the notation for the resonance half-width (\ref{Width}), we arrive at expression:

\begin{equation}
    M_z = \frac{\Delta(R_c - R_p)}{\Delta^2 + \Omega^2}M_0\,,
    \label{SolutionMz}
\end{equation}

During propagation through the medium, the intensities of the waves ${\bf E}_c$ and ${\bf E}_p$  obey the well-known Beer-Bouguer-Lambert law:

\begin{equation}\label{BugerLaw}
    \frac{dI_{c,p}}{dz} = -\varkappa\,I_{c,p}\,\bigl[1-s\zeta\bigl(M_z/M_0\bigr)\bigr]\,.
\end{equation}

\noindent In this equation, $s$$\,=\,$$\pm1$ depending on the orientation of the spin of the pumping wave photon with respect to the quantization axis $z$ \cite{Shah2009}. In our case, $s = +1$. The spin of the probe wave photon is assumed to be always oriented in the opposite direction to that of the pumping wave photon.

In most research works, the theoretical model describing a GSHE-based sensor assumes a sufficiently high buffer gas pressure, so that absorption in the cell is represented by a single broad profile without good spectral resolution of the ground-state hyperfine structure \cite{Shah2009}. In our case, however, the buffer gas pressure can be considered intermediate in magnitude: the hyperfine structure is unresolved only in the excited state, while in the ground state it is clearly visible, as seen in Fig.~\ref{fig:3}(b). Under such conditions, the laser light selectively excites the levels $F_g$$\,=\,$$3$ and $F_g$$\,=\,$$4$, which is reflected in the observed MOR. In particular, this affects the sign of the resonances. To account for this effect, we introduced into Eq. (\ref{BugerLaw}) an additional parameter $\zeta$, which also takes the value $-1$ or $+1$ depending on the type of dipole transition in the D$_1$ line. Based on a more rigorous analysis using the density matrix formalism, it can be shown that the sign of $\zeta$ is determined by the expression:

\begin{eqnarray}
    {\rm sign}[\zeta] = {\rm sign}\Biggl[(-1)^{F_g+F_e} \SixJ{1}{1}{1}{F_g}{F_g}{F_e}\Biggr]\,,
\end{eqnarray}

\noindent where the curly brackets $\{\dots\}$ denote the Wigner $6j$ symbols \cite{Varshalovich1988}. In magnetometry problems, the most commonly considered transitions are of the type $F_g$$\,=\,$$F$$\to$$F_e$$\,=\,$$F$ and $F_g$$\,=\,$$F$$\to$$F_e$$\,=\,$$F$$-$$1$, so-called ``dark'' transitions, for which $\zeta$$\,=\,$$+1$. These transitions are characterized by the fact that when excited by a pump wave with $\sigma^+$ ($\sigma^-$) polarization, there exists an end Zeeman sub-level of the ground state with $m_g = F$ ($-F$) that does not interact with the resonant radiation (this state is also called ``stretched'' state). As a result, due to optical pumping, a significant fraction of atoms accumulates in this state, leading to transparency of the resonant medium for the pump wave. The presence of a magnetic field prevents optical pumping into the end Zeeman sub-level, which increases absorption in the medium. These processes give rise to the observation of MOR in the form of electromagnetically induced transparency [Fig.~\ref{fig:6}(a), upper curve]. Conversely, the probe wave with $\sigma^-$ ($\sigma^+$) polarization experiences strong absorption when interacting with atoms in the sub-level $m_g$$\,=\,$$-F$ ($F$), leading to the observation of electromagnetically induced absorption in the probe wave intensity during scanning of the magnetic field [Fig.~\ref{fig:6}(a), lower curve].

The sign of MOR is determined not only by the physical effects described above. Thus, with increasing light intensity, atoms are increasingly pumped out of the working energy level into another hyperfine component of the ground state, which does not interact with the resonant light. This is equivalent to a reduction in the degree of atomic spin polarization in Eq. (\ref{BugerLaw}). If the contribution of this process becomes significant, the sign of the observed resonances may change. In particular, for optical transitions from $F_g$$\,=\,$$3$, at higher pump-wave intensity the atoms are transferred to another hyperfine component with $F_g$$\,=\,$$4$. In this case, the medium becomes more transparent both for the pump and for the probe wave, resulting in EIT being observed in both channels of the balanced photodetector [Fig. \ref{fig:6}(d)]. For transitions from $F_g$$\,=\,$$4$, however, at any pump intensity there exists an end Zeeman sub-level that does not interact with the pump wave. Therefore, for such (``dark'') transitions the sign of MOR does not change with increasing pump intensity [Fig. \ref{fig:6}(c)].

Under the assumption of optically thin medium (i.e., $M_z$ depends weakly on $z$), the solution of Eq. (\ref{BugerLaw}) reads:

\begin{eqnarray}
I_{c,p}\,(L) &&= I_{c,p}(0) \, e^{-\varkappa L\,\bigl[1-s\zeta\bigl(M_z/M_0\bigr)\bigr]} \nonumber\\
&& \approx I_{c,p}(0) \,\bigl[1 - \varkappa L + \varkappa L s\, \zeta\,\bigl(M_z/M_0\bigr)\bigr]\,,
\end{eqnarray}

\noindent At $\zeta$$\,=\,$$-1$, $s$$\,=\,$$+1$$\,(-1)$ for pump (probe) wave, and $M_z$ from (\ref{SolutionMz}), the latter expression leads to (\ref{IntensC}) and (\ref{IntensP}) in Section~\ref{sec:2}.


\begin{thebibliography}{67}


\bibitem{Allred2002}
J.~C.~Allred, R.~N.~Lyman, T.~W.~Kornack, \textit{et al.}, High-sensitivity atomic magnetometer unaffected by spin-exchange relaxation, Phys. Rev. Lett. {\bf 89}, 130801 (2002).

\bibitem{Kominis2003}
I.~K.~Kominis, T.~W.~Kornack, J.~C.~Allred, \textit{et al.}, A subfemtotesla multichannel
atomic magnetometer, Nature {\bf 422}, 596 (2003).

\bibitem{Budker2007}
D.~Budker and M.~V.~Romalis, Optical magnetometry, Nat. Phys. {\bf 3}, 227 (2007).


\bibitem{Xiao2023}
W.~Xiao, C.~Sun, L.~Shen, \textit{et al.}, A movable unshielded magnetocardiography system, Sci. Adv. {\bf 9}, eadg1746 (2023).

\bibitem{Chen2025}
J.~Chen, C.~Ye, X.~Hou, \textit{et al.}, Bias calibration of optically pumped magnetometers based on
variable sensitivity, Sensors {\bf 25}, 433 (2025).






\bibitem{Osborne2018}
J.~Osborne, J.~Orton, O.~Alem, \textit{et al.}, Fully integrated standalone zero field optically pumped magnetometer for biomagnetism, SPIE Proc. {\bf 10548}, 105481G (2018).

\bibitem{Boto2022}
E.~Boto, V.~Shah, R.~M.~Hill, \textit{et al.}, Triaxial detection of the neuromagnetic field using optically-pumped magnetometry: feasibility and application in children, NeuroImage {\bf 252}, 119027 (2022).

\bibitem{Alem2023}
O.~Alem, K.~J.~Hughes, I.~Buard, \textit{et al.}, An integrated full-head OPM-MEG system based on 128 zero-field sensors, Front. Neurosci. {\bf 17}, 1190310 (2023).

\bibitem{Fedosov2025}
N.~Fedosov, D.~Medvedeva, O.~Shevtsov, \textit{et al.}, A reliable and reproducible real-time access to sensorimotor rhythm with a small number of optically pumped magnetometers, J. Neural Eng. {\bf 22}, 046031 (2025).

\bibitem{Fabricant2021}
A.~Fabricant, G.~Z.~Iwata, S.~Scherzer, \textit{et al.}, Action potentials induce biomagnetic fields in carnivorous Venus flytrap plants, Sci. Rep. {\bf 11}, 1438 (2021).

\bibitem{Taskova2022}
E.~Taskova, E.~Alipieva, S.~Kolev, \textit{et al.}, Coherent optical spectroscopy characterization of the magnetic properties of oriented Fe$_3$O$_4$ nanoparticles, J. Phys. Conf. Ser. {\bf 2240}, 012022 (2022).

\bibitem{Wickenbrock2013}
A.~Wickenbrock, F.~Tricot, and F.~Renzoni, Magnetic induction measurements using an all-optical $^{87}$Rb atomic magnetometer, Appl. Phys. Lett. {\bf 103}, 243503 (2013).

\bibitem{Afach2018}
S.~Afach, D.~Budker, G.~DeCampet, \textit{et al.}, Characterization of the global network of optical magnetometers to search for exotic physics (GNOME), Phys. Dark Universe {\bf 22}, 162 (2018).


\bibitem{Ellmeier2023}
M.~Ellmeier, C.~Amtmann, A.~Pollinger, \textit{et al.}, Frequency shift compensation for single and dual laser beam pass sensors of a coherent population trapping resonance based coupled dark state magnetometer, Measurement: Sensors {\bf 25}, 100606 (2023).

\bibitem{Aleksandrov2009}
E.~B.~Aleksandrov and A.~K.~Vershovskii, Modern radio-optical methods in quantum magnetometry, Phys. Usp. {\bf 52}, 573 (2009).

\bibitem{Fabricant2023}
A.~Fabricant, I.~Novikova, and G.~Bison, How to build a magnetometer with thermal atomic vapor: a tutorial, New J. Phys. {\bf 25}, 025001 (2023).

\bibitem{Gawlik1991}
W.~Gawlik, D.~Gawlik, and H.~Walther, in The Hanle Effect and Level-Crossing Spectroscopy, The Hanle Effect and Atomic Physics,  G. Moruzzi and F. Strumia, eds. (Springer, New York, 1991) Chapter 2, pp. 47-85.

\bibitem{Breschi2012}
E.~Breschi and A.~Weis, Ground-state Hanle effect based on atomic alignment, Phys. Rev. A {\bf 86}, 053427 (2012).

\bibitem{Alipieva2005}
E.~Alipieva, S.~V.~Gateva, and E.~Taskova, Potential of the single-frequency CPT resonances for magnetic field measurement, IEEE Trans. Instrum. Meas. {\bf 54}(2), 738 (2005).

\bibitem{Shah2007}
V.~Shah, S.~Knappe, P.~D.~D.~Schwindt, \textit{et al.}, Subpicotesla atomic magnetometry with a microfabricated vapour cell, Nat. Photonics \textbf{1}, 649 (2007).

\bibitem{Papoyan2016}
A.~Papoyan, S.~Shmavonyan, A.~Khanbekyan, \textit{et al.}, Magnetic-field-compensation optical vector magnetometer, Appl. Opt. {\bf 55}(4), 892 (2016).

\bibitem{LeGal2019}
G.~Le~Gal, G.~Lieb, F.~Beato, \textit{et al.}, Dual-axis Hanle magnetometer based on atomic alignment with a single optical access, Phys. Rev. Applied {\bf 12}, 064010 (2019).

\bibitem{Dong2012}
H.~F.~Dong, J.~C.~Fang, B.~Q.~Zhou, \textit{et al.}, Three-dimensional atomic magnetometry, Eur. Phys. J. Appl. Phys. {\bf 57}, 21004 (2012).

\bibitem{Azizbekyan2017}
H.~Azizbekyan, S.~Shmavonyan, A.~Khanbekyan, \textit{et al.}, High-speed optical three-axis vector magnetometry based on nonlinear Hanle effect in rubidium vapor, Opt. Eng. {\bf 56}, 074104 (2017).


\bibitem{LeGal2022}
G.~Le~Gal and A. Palacios-Laloy, Zero-field magnetometry based on the combination of atomic orientation and alignment, Phys. Rev. A {\bf 105}, 043114 (2022).

\bibitem{Holmes2022}
N.~Holmes, M.~Rea, J.~Chalmers, \textit{et al.}, A lightweight magnetically shielded room with active shielding, Sci. Rep. {\bf 12}, 13561 (2022).

\bibitem{Skidchenko2025}
E.~Skidchenko, A.~Butorina, N.~Fedosov, \textit{et al.}, The tale of two rooms: comparison of QuSpin zero-field OPMs operation in two magnetically shielded environments, IEEE Tran. Instrum. Meas. {\bf 74}, 9516511 (2025).

\bibitem{Zhang2025}
Y.~Zhang, Z.~Wang, L.~Cao, \textit{et al.}, Enhanced gradient field compensation in multi-channel atomic magnetometers with adaptive algorithms, Adv. Quantum Technol. {\bf 8}, 2400346 (2025).



\bibitem{Happer1977}
W.~Happer and A.~C.~Tam, Effect of rapid spin exchange on the magnetic-resonance spectrum of alkali vapors, Phys. Rev. A {\bf 16}, 1877 (1977).

\bibitem{Zhang2022}
S.~Zhang, J.~Lu, Y.~Zhou, \textit{et al.}, Zero field optically pumped magnetometer with independent dual-mode operation, Chin. Opt. Lett. {\bf 20}, 081202 (2022).

\bibitem{Twinleaf}
``Twinleaf'' LLC, https://twinleaf.com

\bibitem{Shah2009}
V.~Shah and M.~V.~Romalis, Spin-exchange relaxation-free magnetometry using elliptically polarized light, Phys. Rev. A {\bf 80}, 013416 (2009).

\bibitem{Tang2021}
J.~Tang, Y.~Zhai, L.~Cao, \textit{et al.}, High-sensitivity operation of a single-beam atomic magnetometer for three-axis magnetic field measurement, Opt. Express {\bf 29}, 15641 (2021).

\bibitem{Sebbag2021}
Y.~Sebbag, E.~Talker, A.~Naiman, \textit{et al.}, Demonstration of an integrated nanophotonic chip-scale alkali vapor magnetometer using inverse design, Light: Science \& Applications {\bf 10}, 54 (2021).

\bibitem{Yang2023}
X.~Yang, M.~Benelajla, S.~Carpenter, \textit{et al.}, Analysis of atomic magnetometry using metasurface optics for balanced polarimetry, Opt. Express {\bf 31}, 13436 (2023).

\bibitem{Hu2017}
Y.~Hu, Z.~Hu, X.~Liu, \textit{et al.}, Reduction of far off-resonance laser frequency drifts based on the second harmonic of electro-optic modulator detection in the optically pumped magnetometer, Appl. Opt. {\bf 56}(21), 5927 (2017).

\bibitem{Petrenko2021}
M.~V.~Petrenko, A.~S.~Pazgalev, and A.~K.~Vershovskii, Single-beam all-optical nonzero-field magnetometric sensor for magnetoencephalography applications, Phys. Rev. Appl. {\bf 15}, 064072 (2021).

\bibitem{Petrenko2025}
M.~V.~Petrenko and A.~K.~Vershovskii, Anomalous suppression of spin-exchange relaxation in alignment signals in cesium in ultraweak magnetic fields, Phys. Rev. A {\bf 112}, 013123 (2025).

\bibitem{Corvilain2025}
P.~Corvilain, V.~Wens, M.~Bourguignon, {\it et al.}, Pushing the boundaries of MEG based on optically pumped magnetometers towards early human life, Imaging Neurosci. {\bf 3}, imag\_a\_00489 (2025).

\bibitem{Wang2025}
S.~Wang, J.~Lu, K.~Zhang, \textit{et al.}, Zero-field atomic magnetometer to extract longitudinal magnetic field, Phys. Rev. Research {\bf 7}, L032024 (2025).

\bibitem{Petrenko20212}
M.~V.~Petrenko, A.~S.~Pazgalev, and A.~K.~Vershovskii, Ultimate parameters of the all-optical single-beam nonzero magnetic field sensor for biological applications, IEEE Mag. Lett. {\bf 12}, 8104605 (2021).

\bibitem{Rushton2023}
L.~M.~Rushton, L.~Elson, A.~Meraki, \textit{et al.}, Alignment-based optically pumped magnetometer using a buffer-gas cell, Phys. Rev. Appl. {\bf 19}, 064047 (2023).

\bibitem{Bonnet2025}
M.~Bonnet, D.~Schwartz, T.~Gutteling, \textit{et al.}, A fully integrated whole-head helium OPM MEG: a performance assessment compared to cryogenic MEG, Front. Med. Technol. {\bf 7}, 1548260 (2025).

\bibitem{Makarov2025}
A.~Makarov, K.~Kozlova, D.~Brazhnikov, \textit{et al.}, All-optical atomic magnetometry using an elliptically polarized amplitude-modulated light wave, Opt. Commun. {\bf 577}, 131369 (2025).

\bibitem{Krzyzewski2019}
S.~P.~Krzyzewski, A.~R.~Perry, V.~Gerginov, \textit{et al.}, Characterization of noise sources in a microfabricated single-beam zero-field optically-pumped magnetometer, J. Appl. Phys. {\bf 126}, 044504 (2019).

\bibitem{Ma2022}
N.~Ma, L.~Duan, D.~Ma, \textit{et al.}, Demonstration of a high-density alkali-metal atomic magnetometer based on the frequency-symmetrical detuning effect of two pumping lights, Opt. Express {\bf 30}, 45930 (2022).

\bibitem{Peng2024}
J.~Peng, Y.~Yin, A.-N.~Xu, \textit{et al.}, Signal-enhanced high-sensitivity atomic magnetometer based on multi-pass cell, Appl. Phys. Express {\bf 17}, 112003 (2024).

\bibitem{Johnson2010}
C.~Johnson, P.~D.~D.~Schwindt, and M.~Weisend, Magnetoencephalography with a two-color pump-probe, fiber-coupled atomic magnetometer, App. Phys. Lett. {\bf 97}, 243703 (2010).

\bibitem{Zhao2023}
B.~Zhao, J.~Tang, L.~Li, \textit{et al.}, Femtotesla $^{87}$Rb magnetometer based on a coaxial pump-probe beam delivery system, Sensors \& Actuators: A. Physical {\bf 364}, 114856 (2023).

\bibitem{Varshalovich1988}
D.~A.~Varshalovich, A.~N.~Moskalev, and V.~K.~Khersonskii, Quantum Theory of Angular Momentum (World Scientific Publishing, Singapore, 1988).

\bibitem{Dehmelt1957}
H.~G.~Dehmelt, Modulation of a light beam by precessing absorbing atoms, Phys. Rev. {\bf 105}(6), 1924 (1957).

\bibitem{Slocum1973}
R.~E.~Slocum and B.~I.~Marton, Measurement of weak magnetic fields using zero-field parametric resonance in optically pumped He$^4$, IEEE Trans. Magnet. {\bf MAG-9}(3), 221 (1973).

\bibitem{Savukov2005}
I.~M.~Savukov, S.~J.~Seltzer, M.~V.~Romalis, \textit{et al.}, Tunable atomic magnetometer for detection of radio-frequency magnetic fields, Phys. Rev. Lett. {\bf 95}, 063004 (2005).

\bibitem{Brazhnikov2022}
D.~V.~Brazhnikov, V.~I.~Vishnyakov, A.~N.~Goncharov, \textit{et al.}, Level-crossing resonances on open atomic transitions in a buffered Cs vapor cell: linewidth narrowing, high contrast, and atomic magnetometry applications, Phys. Rev. A {\bf 106}, 013113 (2022).

\bibitem{Brazhnikov2021}
D.~V.~Brazhnikov, V.~I.~Vishnyakov, S.~M.~Ignatovich, \textit{et al.}, High-contrast level-crossing resonances in a small cesium vapor cell for applications in atomic magnetometry, Appl. Phys. Lett. \textbf{119}, 024001 (2021).

\bibitem{Brazhnikov2019}
D.~V.~Brazhnikov, S.~M.~Ignatovich, A.~S.~Novokreshchenov, \textit{et al.}, Ultrahigh-quality electromagnetically induced absorption resonances in a cesium vapor cell, J. Phys. B: At. Mol. Opt. Phys. \textbf{52}, 215002 (2019).

\bibitem{Brazhnikov2020}
D.~Brazhnikov, S.~Ignatovich, I.~Mesenzova, \textit{et al.}, Shift of zero-field level-crossing resonance in the Cs D$_1$ line and its use in vector magnetometry, Opt. Lett. {\bf 45}, 3309 (2020).

\bibitem{Makarov2023}
A.~O.~Makarov, D.~V.~Brazhnikov, and A.~N.~Goncharov, Observation of the strong magneto-optical rotation of the polarization of light in rubidium vapor for applications in atomic magnetometry, JETP Lett. {\bf 117}, 509 (2023).

\bibitem{Oelsner2019}
G.~Oelsner, V.~Schultze, R.~IJsselsteijn, \textit{et al.}, Performance analysis of an optically pumped magnetometer in Earth's magnetic field, EPJ Quantum Technology {\bf 6}(1), 1 (2019).

\bibitem{Chen2022}
Y.~Chen, L.~Zhao, N.~Zhang, \textit{et al.}, Single beam Cs-Ne SERF atomic magnetometer with the laser power differential method, Opt. Express {\bf 30}(10), 16541 (2022).

\bibitem{Ma2024}
Y.~Ma, Y.~Chen, M.~Yu, \textit{et al.}, Ultrasensitive SERF atomic magnetometer with a miniaturized hybrid vapor cell, Microsystems \& Nanoengineering {\bf 10}, 121 (2024).

\bibitem{Nienhuis1998}
G.~Nienhuis and F.~Schuller, Magneto-optical effects of saturating light for arbitrary field direction, Opt. Commun. {\bf 151}, 40 (1998).

\bibitem{Popov2018}
E.~N.~Popov, V.~A.~Bobrikova, S.~P.~Voskoboinikov, \textit{et al.}, Features of the formation of the spin polarization of an alkali metal at the resolution of hyperfine sublevels in the $^2$$S$$_{1/2}$ state, JETP Lett. {\bf 108}, 513 (2018).

\bibitem{VitaWave}
``Vitawave'', https://vitawave.ru

\bibitem{Kornack2007}
T.~W.~Kornack, S.~J.~Smullin, S.-K.~Lee, \textit{et al.}, A low-noise ferrite magnetic shield, Appl. Phys. Lett. {\bf 90}, 223501 (2007).

\bibitem{Lu2020}
J.~Lu, D.~Ma, K.~Yang, \textit{et al.}, Study of magnetic noise of a multi-annular ferrite shield, IEEE Access {\bf 8}, 40918 (2020).

\bibitem{Petrenko2022}
M.~Petrenko and A.~Vershovskii, Towards a practical implementation of a single-beam all-optical non-zero-field magnetic sensor for magnetoencephalographic complexes, Sensors {\bf 22}, 9862 (2022).

\bibitem{Martinez2010}
R.~Jim\'enez-Mart\'{\i}nez, W.~Clark~Griffith, Y.-J.~Wang, \textit{et al.}, Sensitivity comparison of Mx and frequency-modulated Bell-Bloom Cs magnetometers in a microfabricated cell, IEEE Trans. Instrum. Meas. {\bf 59}, 372 (2010).

\bibitem{Zhao2024}
B.~Zhao, J.~Tang, L.~Li, \textit{et al.}, Transient dynamics of magneto-optic rotation with elliptically polarized light, Results in Physics {\bf 60}, 107686 (2024).

\bibitem{Cohen1970}
C.~Cohen-Tannoudji, J.~Dupont-Roc, S.~Haroche, \textit{et al.}, Diverses r\'esonances de croisement de niveaux sur des atomes pomp\'es optiquement en champ nul. I. Th\'eorie, Rev. Phys. Appl. (Paris) {\bf 5}, 95 (1970).

\bibitem{Li2023}
X.~Li, Z.~Guo, R.~Yang, \textit{et al.}, Single-beam double-pass miniaturized atomic magnetometer for biomagnetic imaging systems, IEEE Sensors J. {\bf 23}, 12433 (2023).


\bibitem{Vershovskii2020}
A.~K.~Vershovskii, S.~P.~Dmitriev, G.~G.~Kozlov, \textit{et al.}, Projection spin noise in optical quantum sensors based on thermal atoms, Tech. Phys. {\bf 65}, 1193 (2020).

\bibitem{Budker2023}
D.~Budker and M.~G.~Kozlov, Sensing: equation one, Opt. Mem. Neural Networks {\bf 32}, S409 (2023).

\bibitem{Klassen2024}
W.~Klassen, S.~Ahmed, K.~P.~Grehan, \textit{et al.}, Demonstration of magnetically silent optically pumped magnetometers for the TUCAN electric dipole moment experiment, Eur. Phys. J. C {\bf 84}, 1181 (2024).

\bibitem{Horrom2012}
T.~Horrom, R.~Singh, J.~P.~Dowling, \textit{et al.}, Quantum-enhanced magnetometer with low-frequency squeezing, Phys. Rev. A {\bf 86}, 023803 (2012).

\bibitem{Li2022}
J.~Li and I.~Novikova, Improving sensitivity of an amplitude-modulated magneto-optical atomic magnetometer using squeezed light, J. Opt. Soc. Am. B {\bf 39}, 2998 (2022).

\bibitem{Lee2008}
S.-K.~Lee and M.~V.~Romalis, Calculation of magnetic field noise from high-permeability magnetic shields and conducting objects with simple geometry, J. Appl. Phys. {\bf 103}, 084904 (2008).







\end{thebibliography}
\end{document}